\documentclass[manuscript]{acmart}

\AtBeginDocument{%
  }

\acmConference[]{}
\acmBooktitle{}
\acmDOI{}
\acmISBN{}
\acmPrice{}

\renewcommand\footnotetextcopyrightpermission[1]{} % removes footnote with conference information
\setcopyright{none}
\renewcommand\footnotetextcopyrightpermission[1]{}
\usepackage{wrapfig}

\acmSubmissionID{1389}

\begin{document}

\title{HIDAgent: A Toolkit Enabling ``Personal Agents'' on HID-Compatible Devices}

%\author{Submission \#1389}
\author{Jeffrey P. Bigham}
\email{jbigham@cs.cmu.edu}
\affiliation{%
  \institution{Carnegie Mellon University}
  \city{Pittsburgh}
  \state{PA}
  \country{USA}
}

%\author{Submission \#1389}

\begin{abstract}

UI Agents powered by increasingly performant AI promise to eventually use computers the way that people do – by visually interpreting UIs on screen and issuing appropriate actions to control them ({\em e.g.}, mouse clicks and keyboard entry). While significant progress has been made on interpreting visual UIs computationally, and in sequencing together steps to complete tasks, controlling UIs is still done with system-specific APIs or VNC connections, which limits the platforms and use cases that can be explored. This paper introduces HIDAgent, an open-source hardware/software toolkit enabling UI agents to operate HID-compatible computing systems by emulating the physical keyboard and mouse. HIDAgent is built using three off-the-shelf components costing less than \$30 and a Python library supporting flexible integration. We validated the HIDAgent toolkit by building five diverse use case prototypes across mobile and desktop platforms. As a hardware device, HIDAgent supports research into new interaction scenarios where the agents are separated from the devices they control.
\end{abstract}

%\begin{CCSXML}
%<ccs2012>
%
%\end{CCSXML}

%\ccsdesc[500]{Do Not Use This Code~Generate the Correct Terms for Your Paper}

%\keywords{Do, Not, Us, This, Code, Put, the, Correct, Terms, for, Your, Paper}
\begin{teaserfigure}
  \includegraphics[width=\textwidth]{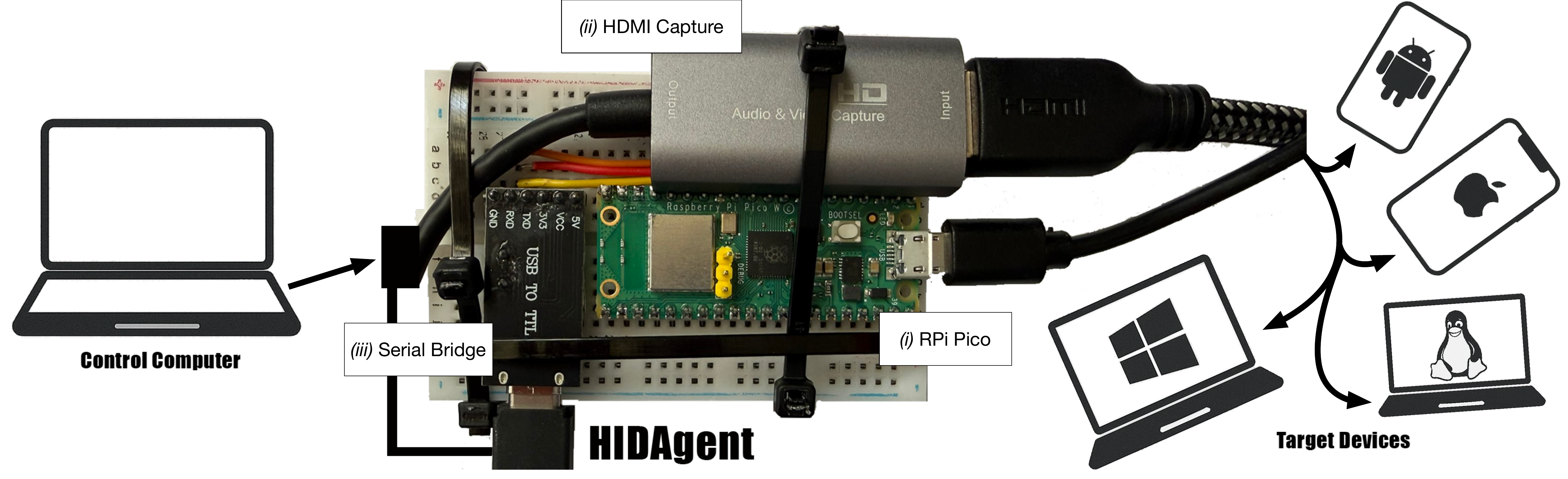}
  \caption{The HIDAgent toolkit is composed of three off-the-self hardware components ({\em (i)} a RP2040 microcontroller, {\em (ii)} an HDMI to USB converter, and {\em (iii)} a CH340 USB-to-Serial bridge) and a Python software library that makes it easy to develop programs running on a control computer that observe and operate target HID-compatible devices like destops and mobile phones.}
  \Description{}
  \label{fig:hidagent}
\end{teaserfigure}

\maketitle

\section{Introduction}

%UI Agents powered by increasingly performant AI promise to eventually use computers the way that people are assumed to do – by visually interpreting UIs on screen and issuing appropriate actions to control them ({\em e.g.}, mouse clicks and keyboard entry) \cite{wang2025guiagentsfoundationmodels}.

UI Agents have become increasingly capable \cite{mind2web,amex,agentnet,osgenesis}, with multiple research prototypes and products promising to complete arbitrary computer tasks on a user's behalf ({\em e.g.}, OpenAI Operator \cite{openai_operator}, Anthropic Computer Use \cite{anthropic_2024}, Vercept Vy \cite{verceptvy}, Taxy AI \cite{taxy_ai_2024}).
While end-to-end task completion is far from perfect \cite{xu2025theagentcompanybenchmarkingllmagents}, performance is impressive and continues to improve. UI Agents have also progressed toward using computers more like people do – by visually interpreting graphical user interfaces (GUIs) on screen and issuing appropriate actions to control them ({\em e.g.}, mouse clicks and keyboard entry) \cite{wang2025guiagentsfoundationmodels}. Interestingly, even as UI Agents have improved, the model for how people are assumed to interact with these agents has not changed all that much.

In this paper we explore an alternative model that we call, ``Personal Agents", in which the agent code is separated from the computing system that it controls. Today, most UI Agents must be installed on a target computer (or connected to a remote computer using VNC) \cite{mobileagents3,qin2025ui} and generally perform provided tasks from start to finish. For example, a user can enter a request into Taxy AI, {\em e.g.}, ``Find the cheapest flight from Denver to Paris over the winter holiday,'' and the agent will do its best to operate the web browser to achieve that goal. Recent work has started to challenge some of these assumptions. For example, Morae recognizes when the user might want to make a choice and asks them for their preference at that point \cite{morae}, instead of forcing end-to-end execution.
%
%
%
%In this paper, we challenge the notion that UI Agents must be subservient to a single host system, and instead explore ``personal'' agents that the user can take with them to different platforms.
%HIDAgent, introduced in this paper, works differently from past UI agent stacks because it is separate hardware that
%
HIDAgent, introduced in this paper, works differently than existing UI agent stacks because it is a separate hardware device that operates any device that it is physically connected to.
%. This ``plug-n-play'' architecture means HIDAgent naturally works across devices, works on a user's real setup, and is user controlled.

% These capabilities are unique to HIDAgent and allow for new research into how people might interact with personal UI Agents.

Moving UI Agents to a separate hardware device allows
%``plug-and-play'' 
research into underexplored interaction scenarios, such as,
\vspace{-.5pc}
\begin{itemize}
\item scenarios in which a user wants to or benefits from keeping their agent separate from the system being operated ({\em e.g.}, when users don't trust the target computer, or in “bring your own agent” interactions);
\item scenarios in which the target system is locked down such that custom agent software cannot be installed or installing software might be a hassle ({\em e.g.}, UI agent user studies on participants’ personal devices, platforms that would otherwise prevent agents other than its preferred agent from running); and,
\item scenarios that benefit from the same agent being able to control multiple different kinds of computers ({\em e.g.}, share an agent and its memory between your Android phone and your Windows laptop) and/or for control to be easily moved from one to the other ({\em e.g.}, start a task on an iPhone and complete it on a Linux desktop)
\end{itemize}

HIDAgent is an open-source external hardware/software toolkit that operates HID-compatible {\em target} computing system by directly receiving the pixels of the screen via HDMI capture and then sending appropriate keyboard and mouse commands directly as Human Interface Device (HID) inputs. HIDAgent is built using three off-the-shelf components costing less than \$30 and a supporting Python library. HIDAgent's novel quality is that it allows UI Agents created using it to operate mobile phones (iOS, Android, etc), desktop computers (Mac, Windows, Linux) and even other kinds of devices ({\em e.g.}, virtual reality headsets) just by plugging its USB-C cable into a different device. The HIDAgent code and instructions for putting together the hardware components are released as open source\footnote{https://github.com/CMUBigLab/hidagent} so that other researchers can use them to investigate new ways of interacting with UI agents.

\begin{comment}
These interaction scenarios are not well-supported by existing frameworks 
Existing platforms limit the kinds of research that can be done in this area because they restrict logging of user interactions, cannot be modified to change the presumed interaction, and/or do not easily allow the agents to run on a user's real device with their apps and content. Some platforms, such as Apple iOS and iPad, prevent agent-based research entirely because the platforms are inaccessible for security reasons. We believe these limitations may be a primary reason that relatively little research has been done challenging the machine learning community's focus on the simple task definition and then execution paradigm for how humans will interact with UI agents.
\end{comment}

The primary contribution of this paper is thus the HIDAgent toolkit, which allows a control computer to capture the pixels of the screen from target devices and issues HID commands to operate them. Toolkit contributions in HCI "are evaluated in a holistic fashion according to what they make possible and how they do so" \cite{contribution-types}. Accordingly, to validate HIDAgent, we built five use case prototypes using it, each of which demonstrates different aspects of what is made possible by separating agent control into an external hardware device. As examples, our ``Helpful Observer'' prototype watches the screen until certain criteria is met that could make it useful for the agent to intervene -- we explored this in a ``bring your own agent'' accessibility application, but believe it could be useful as a way to explore agent assistance in other contexts. Our "Extensible UI Agent" prototype allows existing UI agents to control any device, enabling researchers to, {\em e.g.}, conduct studies on the personal devices of participants or conduct studies beyond the artificial environments of current UI agent stacks \cite{mobileagents3,qin2025ui}.

\section{Related Work}

%HIDAgent builds from prior work on {\em (i)} UI frameworks and toolkits, {\em (ii)} understanding UIs visually from their pixels, and {\em (iii)} UI automation and UI Agents.

{\em UI Frameworks and Toolkits} that control and computationally represent graphical user interfaces (GUIs) were created soon after the first GUIs \cite{bradbook}. Simulating human input device (HID) events ({\em e.g.}, mouse movements, keyboard events, etc) with software is well-supported on modern platforms, often via accessibility APIs, although usually requires special permissions on the host computer. Software such as the PyAutoGUI Python library \cite{pyautogui} unifies support for keyboard and mouse simulation across desktop platforms (Mac, Linux, Windows). Some platforms, {\em e.g.}, iOS, do not allow third party use of their APIs. Instead of simulating input devices in software, HIDAgent operates as an external HID device, appearing to the target computer as a regular mouse and keyboard, which means no software needs to be installed. As a result, HIDAgent can control any device that can be operated with a keyboard and mouse (including mobile devices).

The components of early GUIs were simply drawn on the screen and custom code connected received input to intended functions \cite{bradbook} ({\em e.g.}, a program would keep track of where it drew various components on the screen and upon receiving a mouse down event look up what function based on what it previously drew there). This was cumbersome, and so platforms introduced UI toolkits to make development easier. UI toolkits define abstractions like widgets ({\em e.g.}, button, textbox), and define important attributes of them, such as their state ({\em e.g.}, pressed, on, disabled, etc.) and current value ({\em e.g.}, of a text field). The semantic representations of UI elements are then made available via platform-level services for access and control, such as via the Document Objet Model (DOM) on the web and via various APIs on other platforms. These APIs form the foundation of most accessibility and automation approaches \cite{systemclass}. Critically, these APIs only work reliably when developers have provided complete and correct metadata, and nearly all applications across all platforms lack sufficient metadata. More than 90\% of iOS screens \cite{screenrecognition}, Android screens \cite{android-errors}, and web pages \cite{webaimmillion} have been found to contain at least one metadata error, and many contain 10s or 100s of errors.

As a result, {\em Understanding User Interfaces from Pixels} has become a common approach for reliable access to GUIs as computer vision has improved. Early work in this space built from predefined visual templates \cite{prefab,sikuli}. Screen Recognition was one of the first comprehensive approaches for identifying all elements in a GUI screenshot using computer vision \cite{screenrecognition} (released in iOS 14 VoiceOver). A related problem is UI element grounding, which is the connection of a natural language description to a specific coordinate in a screenshot image ({\em e.g.,} "the hamburger menu" $->$ (800,100)), captured in popular benchmarks like ScreenSpot \cite{screenspot} and ScreenSpot-Pro \cite{screenspot-pro}. Visual language models (VLMs) have shown continued improvement on UI grounding \cite{deitke2024molmopixmoopenweights,lu2024omniparser,you2024ferret,qwen3}, with top performing models like Qwen3-VL 30B-A3B nearly saturating the ScreenSpot benchmark (94.5) \cite{qwen3}. HIDAgent only has access to the pixels of the screen, but operating on the pixels instead of the unreliable semantic representations has started to work well.

{\em UI Automation and UI Agents} have a long history in HCI. Early macro systems could repeat previously demonstrated commands exactly, with a variety of attempts to make them more general than the specific captured demonstration \cite{coscripter,sugilite}, often with the goal of being robust to slight user interface or task changes. Many UI Agents operate on the web because the Document Object Model (DOM) of rendered web pages is especially convenient for computational inspection and manipulation. Many UI agents were built for the 2000s web \cite{plow,coscripter,trailblazer,highlight}, since it was powerful enough to capture many different tasks, but also web pages were simple enough that the resulting DOM was simple enough to be reasoned about. Automatic understanding of GUIs from their pixels makes a similarly reliable and simple GUI representation ubiquitous.

Several benchmarks have been released for UI agent task completion \cite{mind2web,amex,agentnet,osgenesis}, which generally evaluate an agent's ability to successfully complete a given task based on a provided description, {\em e.g.}, ``update the cells in the spreadsheet labeled `CHI 2026 revision' with the status DONE''. 
%Benchmarks vary in task complexity and in which platforms are supported (web, desktop, or Android).
Although performance continues to improve, UI agents still fail to reliably complete benchmark tasks of even moderate complexity \cite{osgenesis}. Recent cross-platform benchmark environments, such as Mobile-Agent-v3 \cite{mobileagents3} and UI-TARS \cite{qin2025ui}, do so with either target virtual machines or specialized software that must be installed, unlike HIDAgent. As a result, HIDAgent may make development of future cross-platform benchmarks easier (especially on local platforms and those currently unsupported, like iPhones). More importantly, it may encourage benchmarks for new interaction scenarios supported by HIDAgent.

\section{HIDAgent}

The HIDAgent toolkit (Figure \ref{fig:hidagent}) allows target devices to be observed and operated from a control computer. HIDAgent is composed of both hardware sufficient for capturing screenshots from devices with USB-C/HDMI video output and for performing HID input actions. The included Python library supports using this hardware to create a wide variety of different agents (Section \ref{sec:usecases}). The HIDAgent toolkit hardware is made up of three off-the-self hardware components: {\em (i)} a Raspberry Pi Pico (RP2040) microcontroller, {\em (ii)} an HDMI to
USB converter, and {\em (iii)} a CH340 USB-to-Serial bridge).

\subsection{Simulating HID Events}

Most computers cannot directly serve as HID output devices because their USB chips are limited to receiving HID input, but many low-cost microcontrollers have this capability. We used a Raspberry Pi Pico (RP2040)\footnote{https://www.raspberrypi.com/products/raspberry-pi-pico/} to programmatically simulate HID events, which currently costs ~\$4 USD. The RP2040 only includes one USB-C port, which is used as the HID output device connected to the target computing system. We therefore needed another way to connect the control computer to the RP2040. While the RP Pico W includes Wifi connectivity, a Wifi connection between the Control computer and the RP Pico is undesirable because it requires that both devices to be connected to the same Wifi. Instead, we used the built-in Universal Asynchronous Receiver/Transmitter (UART) functionality on the RP Pico to receive serial commands from the control computer. Many modern laptops also cannot do this directly, and so we used a low-cost USB-C to TTL UART serial bridge. For this, we chose the CH340\footnote{https://cdn.sparkfun.com/datasheets/Dev/Arduino/Other/CH340DS1.PDF}, which currently costs \$6 USD, although there are other comparable options.

We wrote custom controller code for the RP2040 in CircuitPython\footnote{https://circuitpython.org/}, which provides a limited version of the Python language able to run on the RP Pico microcontroller. The primary functions of this code are to {\em (i)} receive the serial commands over UART sent from the control computer and reply back to indicate the commands were sent, and {\em (ii)} send the HID commands to the target device. Commands and responses are sent via JSON, {\em e.g.}:

\begin{verbatim}
send: {"type": "click", "x": 121, "y": 2145}
receive: {"result": "success"}
\end{verbatim}

One challenge we ran into initially was that the target devices would get confused if the HID commands were sent too quickly, since humans do not interact so quickly, and so the RP2040 code includes appropriate delays built in, implemented via short sleep commands of, {\em e.g.}, 0.1 seconds. Because these delays are only introduced between commands, they do not introduce noticeable latency. For convenience, we also included several keyboard shortcuts we found to be useful as special commands, {\em e.g.}, CMD/Windows+Spacebar named ``run'', which is a useful tool to provide for straightforwardly running applications on Mac, Windows, and iOS.

\subsection{HIDAgent.py}

The UI Agent control computer uses HIDAgent.py to receive and process visual screenshots from the target computer, and to send HID events to the target computer. It provides the following core functions:

\begin{itemize}
    \item \textbf{get\_screenshot()} - identify the connected USB device that is most likely the HDMI capture card, capture one frame, and return it as a Python image object

    \item \textbf{move\_mouse(x,y)} - move the mouse to location (x,y) on the target computer

    \item \textbf{click\_mouse(x,y,<left|right>)} - move the mouse to location (x,y) on the target computer and click the left or right mouse button

    \item \textbf{type(<string>)} - send a string of characters to the target computer

    \item \textbf{keypress(list[keys])} - performs a single keypress (keydown + keyup) of the provided array of keyboard keys on the target computer, which can be provided either as characters or string denoting control keys ({\em e.g.}, ctrl, alt, cmd, space).
\end{itemize}

HIDAgent.py also provides several convenience functions that we found especially useful in our development of use case prototypes that we believe are broadly useful (Section \ref{sec:usecases}):

\begin{itemize}
    \item \textbf{recognize\_gui\_elements(screenshot\_img)} - takes a screenshot image of a GUI as input, and outputs a JSON list of recognize user interface elements, their location, their type, and any content associated with them ({\em e.g.}, a label). Currently, this is implemented as a lightweight wrapper around Omniparser \cite{lu2024omniparser}.
    \item \textbf{llm\_screenshot\_query(screenshot\_img, query, optional:<model>)} - takes a screenshot image of a GUI and a query to run on it, and then outputs a response in JSON format. The default option currently first runs the screenshot through  \texttt{recognize\_gui\_elements} and passes the results and image encoded in base64 to the Gemma:27b model \cite{gemmateam2025gemma3technicalreport}, allowing all of the processing to run locally. We have also built support for OpenAI GPT5 and Anthropic Computer Use (Section \ref{sec:usecases}).
    \item \textbf{run\_application(application\_name)} - exposes ability to run a provided application into one command in a way that works across multiple platforms, {\em i.e.}, the following three commands in sequence (cmd/windows key + spacebar, <application name>, enter)

    \item \textbf{gui\_diff(screenshot1, screenshot2)} - takes in two screenshot images that are the same resolution and returns the location and size of the difference.

    \item \textbf{patch\_location(patch\_img, screenshot)} - takes in an image of patch and returns its location on the provided screenshot or None if it is not detected. This provides an easy mechanism for specific scripting when you know exactly what you want the agent to click on, similar to Sikuli \cite{sikuli}.
    
\end{itemize}

To support debugging, HIDAgent.py provides an optional web-based logging interface on localhost, implemented with Python's http.server, which we found helpful to keep track of GUI states and commands sent, since they can go by quickly on the target device. Each entry in the timestamped visual log contains the captured screenshot, a graphical mark on the screenshot of any command performed on that screen, and a textual log of the command issued and other customizable metadata. HIDAgent also supports sending commands manually to the target device using this interface: clicks on the displayed screenshot are forwarded on to the target device, along with any captured keyboard events.

\subsection{Calibration \& Settings}

\begin{figure}
  \includegraphics[width=\textwidth]{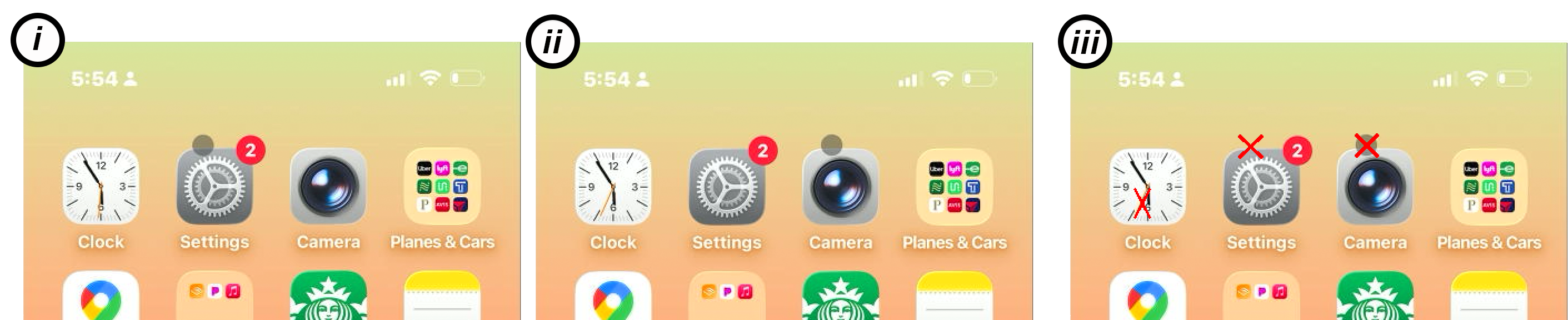}
  \caption{HID space to pixel space calibration is done by observing the cursor's movement in response to HID mouse move events. In this example, the quite faint cursor of the Apple iPhone screen is shown before {\em (i)} and after {\em (ii)}. The pixel locations that were automatically detected to have changed are shown in the third image marked with red x's {\em (iii)}, which includes pixel changes detected in the clock icon because the seconds hand has moved. To differentiate which changes are important, HIDAgent moves the cursor only in one direction at a time and looks for changes only in that direction. In this case, left to right with no vertical movement.}
  \Description{}
  \label{fig:calibration}
\end{figure}

HID devices for pointer/mouse input operate relative to their current position and are interpreted flexibly by different operating systems which may apply different scaling and acceleration to them. As a result, we cannot simply tell HIDAgent to put the mouse pointer at a certain coordinate ({\em e.g.}, (200, 300)) without calibration. We developed a calibration procedure and set of reliable heuristics for getting the mouse pointer to a specified position. Because HIDAgent captures screenshots using HDMI, we always capture the screenshots at 1920x1080 resolution. We remove any black pixels framing the screenshot because other resolutions show up with these ({\em e.g.}, an iPhone in portrait mode). HIDAgent performs an automated calibration procedure when it is first connected in which it moves the mouse to a particular coordinate (100 x 100), takes a screenshot, then moves it to another coordinate (200 x 100), takes another screen shot, and finally computes the difference between recognized changes in the image (Figure \ref{fig:calibration}). One point would be sufficient to compute the mapping between screenshot image coordinates and HID coordinates, but by using two points a known distance apart the system is resilient to other UI changes. To start the mouse from a known coordinate, we first move it to the origin coordinate (0, 0) by issuing several commands to move left and up by large amounts (1000 pixels at a time). This results in the mouse ending at the origin point because the operating system clips values less than zero to avoid the mouse going off screen. To avoid acceleration effects implemented for mouse movements in most operating systems that would otherwise make our pixel to HID space mapping non-linear, we break down large mouse movements into smaller ones of the same size (10px) and issue them repeatedly without delays between.

\subsection{Platform Considerations}

HIDAgent was built to be cross-platform because it emulates the mouse and keyboard, which can be used to control a large number of different platforms. Many common platforms work seamlessly, although we did notice some areas where users would need to set appropriate settings or permissions to allow it work. iOS required the most set up -- we needed to enable both Assistive Touch and Allow Full Keyboard access in the accessibility settings to allow for mouse and keyboard control, although video capture via HDMI worked without setup. Android required us to ``Allow Video Mirroring'' via a dialog that appeared when we connected HIDAgent before it would work. Both Mac and Windows displayed keyboard configuration windows when connecting HIDAgent, but both could be dismissed or ignored.

\section{Use Case Prototypes}
\label{sec:usecases}

To validate the HIDAgent toolkit we created five distinct prototypes using it.
Each prototype is described below and the source code for each is also released as open source (and in supplementary material for submission). The goal of these prototypes is to show how the capabilities of the HIDAgent toolkit allow for flexible exploration of UI agents, not necessarily to demonstrate how well each works; our evaluation is limited to a descriptive analysis of our observations.

\begin{itemize}

\item {\bf Extensible UI Agent}

We implemented two generic UI Agents that can be extended by other researchers. The first connected HIDAgent functions to Claude in Computer Use mode. This was a straightforward adaptation of the standard boilerplate code given for implementing the Computer Use Tool\footnote{https://docs.anthropic.com/en/docs/agents-and-tools/tool-use/computer-use-tool}, which requires handling the following actions when requests by the agent: screenshot, left\_click, type string, key combinations, mouse\_move, and wait. We also added a new runcommand tool that connects to HIDAgent's spotlight command. The second agent uses entirely on-device processing by using Omniparser \cite{lu2024omniparser} to process screenshots and then using Gemma:27b \cite{gemmateam2025gemma3technicalreport} to process the results and the image.

We informally tested both agents with four different operating systems (Android, iOS, Mac, and Windows) on several common UI Agent tasks, {\em e.g.}, "open a new note titled `todo list' for CHI 2026 paper'" and "find the instructions for submitting a CHI 2026 paper". Both were able to suggest and complete some steps of the tasks we gave them, although the Claude model was substantially better. It was much more difficult for both agents to open the correct application on Android since there isn't a dedicated keyboard shortcut for doing so, and so the agents needed to click around to find the right application. Both agents struggled significantly on large resolution desktop computers as compared to mobile phones (a known limitation that is getting better \cite{screenspot-pro}). We tried reducing the resolution of the desktop screens to below 1000px in each dimension as suggested by Claude, but either way the accuracy of locating items on screen was poor. We found that explicitly advising the agent to consider which platform the screenshot was from before suggesting an action to perform helped it avoid silly errors like the agent trying to open Firefox on iOS. This suggests that these models may not be evaluated regularly across the platforms we tried (unsurprising since there are no agent benchmarks for iOS, for instance).

\begin{figure}
  \begin{center}
    \hspace{0pc}{\includegraphics[width=15.5pc]{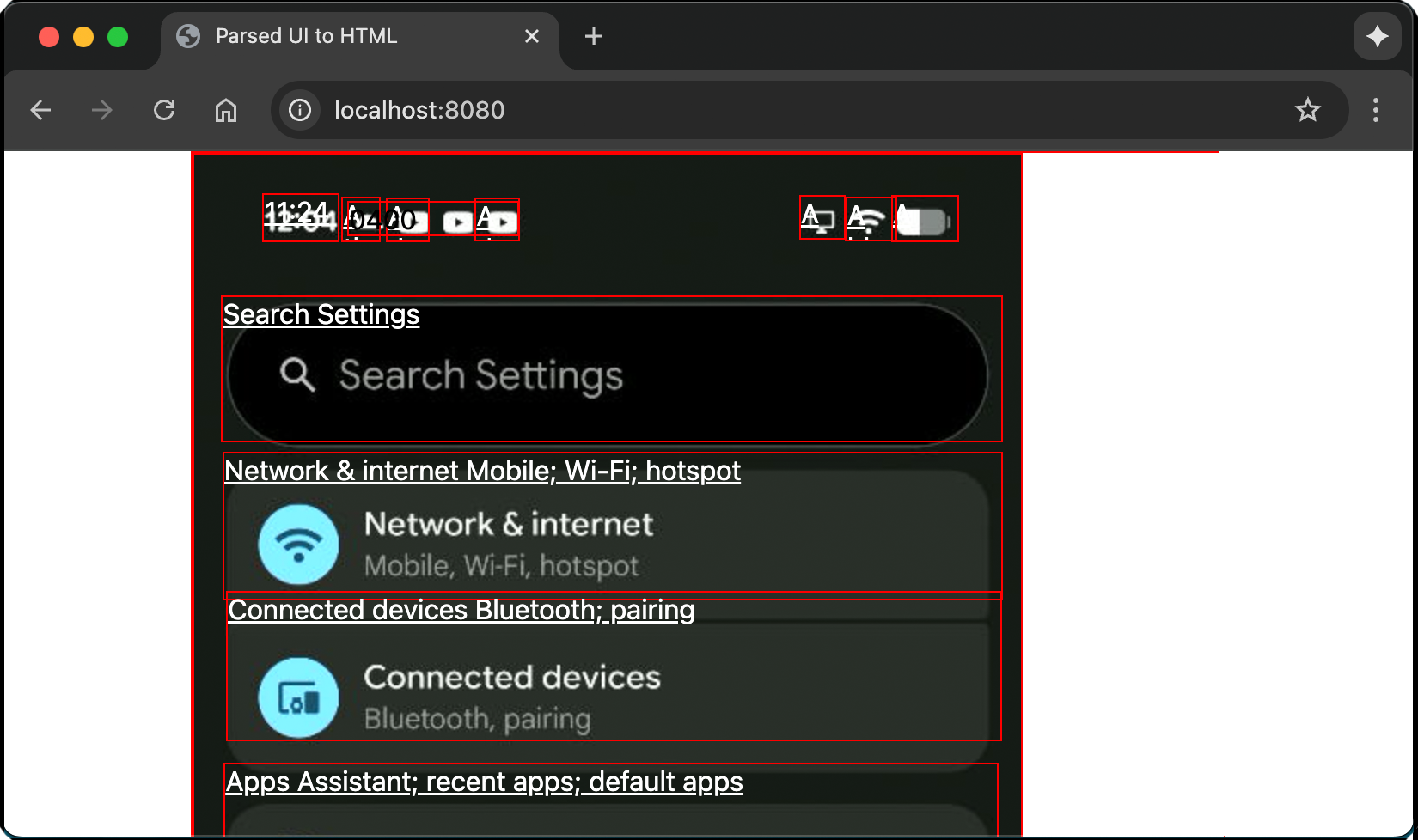}}
  \end{center}
  \caption{This screenshot shows how the Screen Reader Anywhere prototype brings a screenshot from in this case an Android target device into a web page to make it accessible. Content from the pixels of the settings screen is made readable in the HTML and actions performed on the HTML ({\em e.g.}, buttons clicked) are transferred back to the original interface via HIDAgent, enabling users to use their preferred screen reader on the control computer.}
\label{fig:screenreader}
\end{figure}

\item \noindent {\bf Universal UI Data Collection}

Machine learning performance is driven by datasets, and yet there are relatively few datasets that are specific to visual task completion with UI Agents, likely as a result of it being difficult to set up data collection. The Rico dataset of UI screens had crowd workers interact with Android devices over a remote desktop setup \cite{rico}. iOS datasets like the one collected by workers for Screen Recognition \cite{screenrecognition} and automatically in the Never-Ending-Learning of UIs dataset \cite{never-ending} were not released publicly.

Our Universal UI Data Collection prototype uses HIDAgent to create an automated data collection pipeline \cite{never-ending}. The agent starts at whatever screen it is on, stores a screenshot of the screen along with parse results from Omniparser, and then randomly clicks on the screen in different places to see if the screen changes. If the screen changes sufficiently according to gui\_diff, then it is assumed to be a new screen and it repeats the process. This data collection uses the captured screenshot from the HDMI capture, which is lower resolution than the original screen and often includes non-UI visual elements like the mouse pointer. One extension that we explored was using platform-specific screenshot functionality to capture the screens instead. For instance, for Mac a screenshot can be taken using the control sequence cmd+shift+3. This allows full resolution screenshots to be taken without extraneous visual artifacts, but is platform-specific and the data is stored on the target computer. On iOS we can directly use the Assistive Touch nub's menu to select the screenshot function, which is a good use case for the patch\_location function.

%One limitation of this prototype for data collection is that it doesn't directly have access to the underlying view hierarchy or other privileged user interface information from the application.

%\textbf{Observations:}

    \noindent \item {\bf Screen Reader Anywhere}

Similar to UI agents, applications supporting accessibility like screen readers used by people who are blind require significant operating system permissions because they need to interpret running applications and send user interface commands generated by alternative means. For instance, a screen reader may allow an application originally developed for visual use to have its content and state read aloud, and for all functions to be accessed via the keyboard. Fully supporting accessibility requires significant integration and careful coding across the operating system, accessibility  application, and each application that is being used \cite{systemclass}. Difficulty in getting all of these components to come together across every application has led to increasing calls for intermediate ``bots" that could operate the original interface and then expose a preferred accessible interface to users ({\em e.g.}, Vanderheiden's ``Infobot'' \cite{infobot}). Furthermore, oftentimes a person's preferred accessibility software isn't available on the computers they have access to, such as a public kiosk or even a friend's mobile phone.

Our Screen Reader Anywhere prototype (Figure \ref{fig:screenreader}) continually captures a screenshot of the target device, parses through it with Omniparser to identify user interface elements, and creates an accessible web page version of those elements on the control computer. Actions taken on the control computer version of the page (clicking and typing text, for now) are captured and sent back to the target computing device. To facilitate this we added a Python web server that serves the new content and listens for HTTP post requests that contain actions that are then forwarded along to the appropriate HIDAgent functions. Our approach has drawbacks of older screen readers that used a periodically refreshed ``off-screen'' model to enable non-visual web browsing, which could occasionally go stale. Nevertheless, it makes possible non-visual access target devices using a screen reader that wasn't built for them running on the control computer.

\item {\bf Cross-Device Interaction}

HCI researchers have long studied how to enable interactions that go across the many interactive computing devices we have access to at any given moment \cite{crossdevice}. HIDAgent can be easily moved from one device to another, making possible novel cross-device agent interactions in which context is stored on the control computer even as the target computer changes to a new one. This was the simplest prototype to make because it simply involved using the UI Agent prototype previously described in a new way -- as an example, with HIDAgent plugged into a laptop, we asked the agent to find how to enable high contrast mode on an iPhone and then enable it. The idea was that it could start with a web page, and find the information. We then unplugged HIDAgent from the laptop and plugged it into an iPhone where it could complete the interaction.
\vspace{1pc}

\item {\bf The Helpful Observer}

The Helpful Observer prototype allows users to ask for help related to their current screen by accepting input, taking a screenshot, and then offering potential advice. For instance, ``how much is this flight in US dollars" when looking at possible trips on an airline app. The tips can be generated by either Claude or Gemma:27b running locally. Past studies have shown that agents can sometimes be better at figuring out how to navigate confusing or complex user interfaces \cite{cowpilot}. The potential of this prototype in the context of HIDAgent is that you could potentially bring your agent that has experience with you and your past interactions to a new computer just by plugging HIDAgent in, which could potentially help make unlock where and how future agents are allowed to be used. Our prototype required questions to be typed into the commandline on the control computer, but connecting this to voice input could provide an interesting new channel for improving interaction.

\end{itemize}

\section{Discussion \& Future Work}

HIDAgent is a toolkit that can observe and operate a wide variety of different devices, and thus brings up new opportunities and challenges for UI Agent research. In working with our prototype systems, HIDAgent allowed us to find potential applications of agents where state-of-the-art agent models perform poorly. For example, agents could be much better at understanding appropriate actions to take given the operating system that is being controlled -- this matters for everything from how to most efficiently run commands, to how to take a screenshot. UI Agents are known to work best on lower resolution systems, and so when we plugged HIDAgent into our high-resolution work setup it didn't work that well. Fixing these sorts of problems will be necessary to explore how agents can partner with people in their everyday work.

We have considered a number of features that could be added to HIDAgent to make it even more useful. For instance, it could be useful to explore how to incorporate additional context about the user's system. One idea for this is to build a ``context crawler'' that would explore around a system, perhaps try various ways of interacting with the device, peak into the settings on the device to understand the operating system and what peripherals might be attached, etc.

The HIDAgent control computer we used as part of this work was a Mac laptop, although a wide variety of other computing devices could be used for control. Even a minimal microcontroller could be used for control if the computationally expensive processing is done in the cloud ({\em e.g.}, with a call to an LLM API). For instance, a Raspberry Pi 4 may be powerful enough to get the screenshot images from the HDMI capture device and can also emulate HID events, although getting this to work with the particular devices we had did not seem straightforward. This might allow HIDAgent to run in a truly portable device that could be easily plugged into whatever device is nearby. It would be interesting if a mobile phone itself could be used as the control computer, enabling users to truly bring their personal agent with them to other devices they encounter. While technically feasible, we have not figured out in practice how to accomplish both screenshot capture and communication with the HID emulating chip with an existing mobile phone.

The main negative impacts that we consider are those related to how HIDAgent can apply agentic observation and control to computing systems that a person may not have more extensive security permissions on but does have physical access to. Nothing about HIDAgent directly subverts existing security protections of computers, of course -- the agent would still need to enter a password to access computer content and functionality on password-protected systems -- and many computing systems, {\em e.g.} Macs, require permission to be explicitly granted before a new HID input device can be used at all. However, HIDAgent may allow for automation on systems that are ``locked down'' to prevent automation, such as on school-issued laptops that support the connection of an external keyboard and mouse, and thus potentially make it easier to use generative AI tools on such systems against policy.

\section{Conclusion}

This short paper has introduced the idea of ``Personal Agents'', which are computer use agents that exist separate from the computers they control and which are controlled and travel with the person using them. To facilitate research into personal agents, we introduced HIDAgent, a hardware/software toolkit that enables UI agents to observe and operate HID-compatible devices. Through several prototypes, we demonstrated new interaction scenarios made possible with this toolkit that would be much more difficult to achieve with a purely software agent. We believe HIDAgent can be useful for HCI researchers in continuing to innovate and study how humans can work alongside UI agents.

%\begin{acks}
%\end{acks}

\bibliographystyle{ACM-Reference-Format}
\bibliography{hidagent}

\end{document}